\documentclass[10pt,conference,compsocconf,letterpaper]{IEEEtran}
\ifCLASSINFOpdf
\else
\fi

\usepackage{xspace}

\newcommand{\projectName}{\emph{SD-CPS}\xspace}

\usepackage{balance}  
\usepackage{graphicx} 
\usepackage{times}    
\usepackage[utf8]{inputenc}
\usepackage[usenames, dvipsnames]{color}
\usepackage{balance}
\usepackage{moreverb}
\usepackage{amsmath}
\usepackage[utf8]{inputenc}
\usepackage{url}

\usepackage{listings}
\usepackage{color}
\usepackage{textcomp}
\usepackage[noend]{algpseudocode}
\usepackage{algorithm}
\usepackage{enumerate}
\usepackage{varwidth}
\usepackage{xcolor}

\usepackage{amssymb}

\definecolor{mygreen0}{rgb}{0, 0.75, 0}

\definecolor{myred1}{rgb}{1,0,0}
\definecolor{mygreen1}{rgb}{0, 1, 0}
\definecolor{myblue0}{rgb}{0, 0, 1}

\definecolor{myred2}{rgb}{1,0.5,0.5}
\definecolor{mygreen2}{rgb}{0.5, 1, 0.5}
\definecolor{myblue2}{rgb}{0.5, 0.5, 1}

\definecolor{mygreen}{rgb}{0, 0.25, 0}
\definecolor{myblue}{rgb}{0, 0, 0.75}
\definecolor{myred0}{rgb}{0.5,0,0}

\definecolor{listinggray}{gray}{0.98}
\definecolor{lbcolor}{rgb}{0.98,0.98,0.98}
\begin{document}
%
\bibliographystyle{IEEEtran}
\title{\projectName: Taming the Challenges of Cyber-Physical Systems with a Software-Defined Approach}

\author{\IEEEauthorblockN{Pradeeban Kathiravelu}
\IEEEauthorblockA{INESC-ID Lisboa / Instituto Superior Técnico\\
Universidade de Lisboa\\
Rua Alves Redol 9, Lisboa 1000-029, Portugal\\
Tel: (+351) 21 310 0300\\
Fax: (+351) 21 314 5843\\
pradeeban.kathiravelu@tecnico.ulisboa.pt}
\and
\IEEEauthorblockN{Lu{\'\i}s Veiga}
\IEEEauthorblockA{INESC-ID Lisboa / Instituto Superior Técnico\\
Universidade de Lisboa\\
Rua Alves Redol 9, Lisboa 1000-029, Portugal\\
luis.veiga@inesc-id.pt}}


%


\maketitle

\begin{abstract}
Cyber-Physical Systems (CPS) revolutionize various application domains with integration and interoperability of networking, computing systems, and mechanical devices. Due to its scale and variety, CPS faces a number of challenges and opens up a few research questions in terms of management, fault-tolerance, and scalability. We propose a software-defined approach inspired by Software-Defined Networking (SDN), to address the challenges for a wider CPS adoption. We thus design a middleware architecture for the correct and resilient operation of CPS, to manage and coordinate the interacting devices centrally in the cyberspace whilst not sacrificing the functionality and performance benefits inherent to a distributed execution.

\end{abstract}


%
\IEEEpeerreviewmaketitle



%
\section{Introduction}
\label{sec:intro}
While the Internet of Things (IoT)~\cite{xia2012internet} motivates for a scenario where there are many smart devices that are all connected together and are accessible pervasively in the Internet, reality is still far from this. We do have several networks of things, where the smart devices (or the ``things'') are interconnected to form network of devices, or connected to an existing enterprise network. However, one needs not to have \textit{the Internet of Things} literally, as it is not necessary to connect everything to the Internet, the single unified network of networks. It is not only unnecessary, but also counter-intuitive to have everything connected and open beyond what is necessary, due to security and privacy reasons.

Cyber-Physical Systems (CPS) fix the shortcomings and limitations in the definition of IoT and similar terms, in clearly defining the common larger ground of theories and practice where the physical/mechanical systems intersect and deeply intertwine with the cyber/computer systems~\cite{lee2015past}. While sharing the core architecture with IoT~\cite{atzori2010internet}, CPS is defined as a pure interdisciplinary mechanism, with applications ranging from smart homes~\cite{munir2014depsys} to smart cities~\cite{pacheco2016design}. Though CPS is a term that is coined relatively in recent times, there have been research and implementation efforts on the topic even before the inception of the term~\cite{leitao2016industrial}.

Due to the scale and variety in its implementation and devices, CPS faces a set of challenges in design and practice~\cite{lee2008cyber}, including: i) unpredictability of the execution environments~\cite{lee2007computing}, ii) communication and coordination within the system~\cite{persson2015communication}, iii) security, distributed fault-tolerance, and recovery upon system and network failures~\cite{cardenas2009challenges}, iv) decision making in the large-scale geo-distributed execution environments, v) modelling and designing the systems~\cite{derler2012modeling}, and vi) management and orchestration of the intelligent agents. 

The challenges are imposed from both the core domains of CPS, including networking, distributed systems, or the physical systems involved, as well as the challenges that manifest due to the co-existence and interdependencies of the cyberspace and physical devices in CPS. A unified approach is necessary to address the challenges that prevent or hinder the seamless adoption, applicability, and reusability of the CPS principles and constructs pervasively.

Software-Defined Networking (SDN) offers reusability and management capabilities, among many other improvements, to networks by separating the control layer as a unified controller, away from the distributed network's data forwarding elements. There have been recent researches on leveraging SDN in the implementation of CPS. SDN has been proposed to improve the resilience of multinetworks in CPS~\cite{qin2014asoftware}. SDN has been leveraged to secure the CPS networks through SDN-assisted emulations~\cite{antonioli2015minicps} and improve the resilience~\cite{dong2015software} of CPS. 

We propose to tackle the current and foreseen future challenges of CPS through a middleware architecture following a software-defined approach. We call the proposed approach for CPS, ``Software-Defined Cyber-Physical Systems (SD-CPS)''. Designed as a middleware platform inspired by the logically centralized control offered by the SDN controllers, \projectName aims to tackle the core challenges of CPS as an architectural enhancement.


\section{An Architecture to \\Tame the Challenges of CPS}
\label{sec:arch}



This section looks into the design of the proposed \projectName approach, and how it attempts to tackle the identified core challenges of CPS~\cite{lee2008cyber}, including modelling, incremental building and testing, execution in a sandbox and production environments, scalability, reusability of services through service compositions, fault-tolerance, and resilience.

Current software-defined approaches can broadly be categorized into: i) approaches that extend or use SDN and SDN controllers, and ii) approaches that follow a similar architecture or motivation of SDN while not actually leveraging SDN as it is. \projectName employs a hybrid approach: It leverages the SDN when SDN is already a part of the deployment architecture of the CPS; however it does not make SDN a pre-requisite, to ensure a wider adoption.

\subsection{\projectName Controller APIs}

\begin{figure}[ht]
	\begin{center}
		\resizebox{\columnwidth}{!}{
			\includegraphics[width=\textwidth]{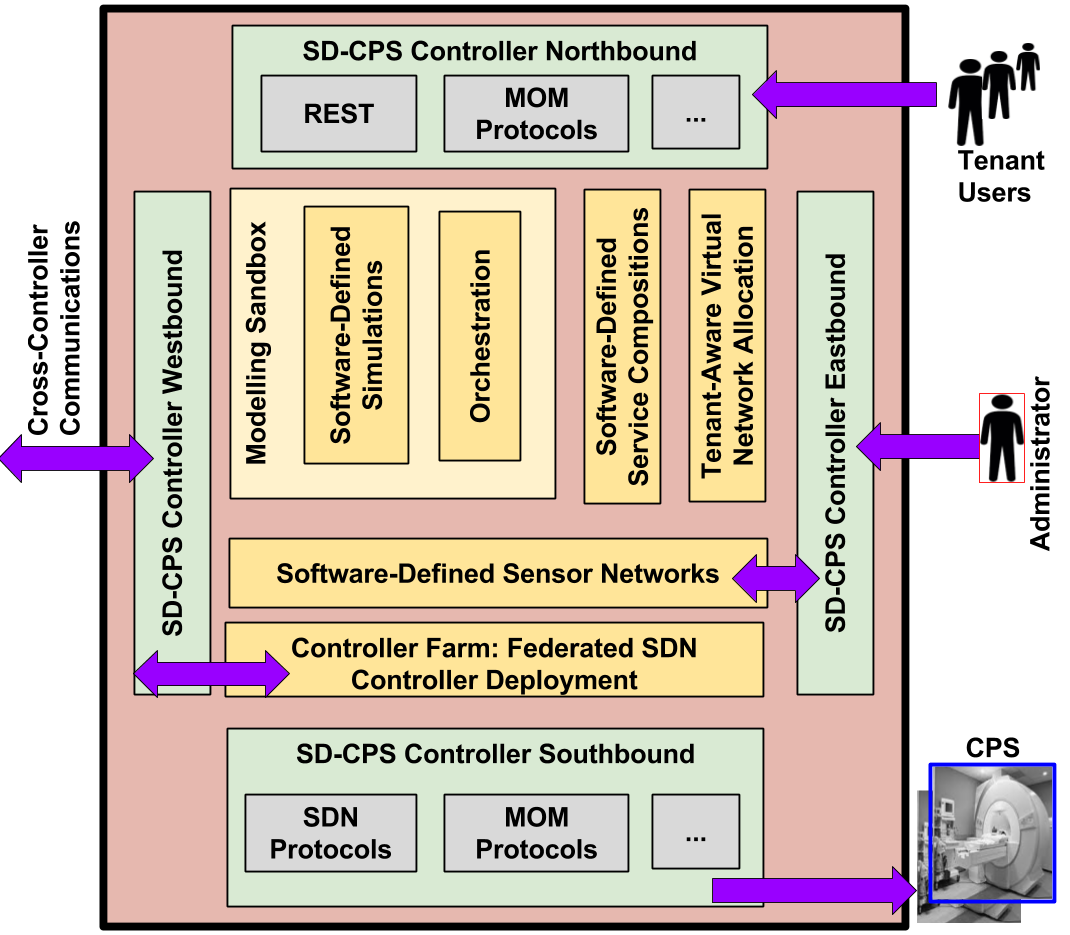}
		}
				\vspace{-2em}
	\end{center}
	\caption{Controller of the ``Cyber'' of \projectName}
				\vspace{-1em}
	\label{fig:bb}
\end{figure}

The core of \projectName is a controller deployment, that controls the ``cyber'' of the CPS and centrally orchestrates the CPS elements. \projectName controller consists of a deployment of multiple SDN controllers and further software components to manage the CPS. The control plane communicates with the underlying network through OpenFlow and other SDN southbound protocol implementations, while communicating with the devices that are non-compliant with OpenFlow through a similar approach inspired by OpenFlow. This ensures that while \projectName has SDN at its core, it is not limited to software-defined networks with SDN switches that are still far from widespread in IoT and CPS settings. \projectName devises its APIs, adapting that of SDN~\cite{jarschel2014interfaces} for the extended distributed controller deployment for CPS. Figure~\ref{fig:bb} depicts the \projectName controller along with the larger \projectName ecosystem. 
				\vspace{-1em}

\textbf{The Northbound API} communicates with the tenant applications, and takes into account the user involvement and human interaction into the CPS. It allows management of smart devices in a tenant-aware manner respecting the tenant intents and system policies as defined from the application layer. It consists of typical SDN northbound protocols including REST and Message-Oriented Middleware (MOM)~\cite{curry2004message} protocols such as Advanced Message Queuing Protocol (AMQP)~\cite{vinoski2006advanced} or MQTT(formerly, Message Queue Telemetry Transport)~\cite{locke2010mq} for the tenant processes to interact with the controller. As MOM protocols are long researched for use with networks of wireless sensors and actuators~\cite{hunkeler2008mqtt,collina2012introducing}, extending SDN with MOM increases its applicability, in addition to scalability.


\textbf{The Southbound API} controls the data plane elements which are physical devices in addition to the regular SDN switches. It handles the communication, coordination, and integration of the network data plane consisting of the CPS devices with the control plane. The southbound consists of typical SDN southbound implementations such as OpenFlow protocol~\cite{mckeown2008openflow} and additional light-weight protocols such as MOM protocols for the communication with the physical devices.

\textbf{The Westbound API} enables inter-control communication among the controllers in \projectName, as well as inter-domain communications across multiple \projectName controller deployments, through their westbound. The controller farm provides a federated deployment of controllers in each \projectName control plane consisting of multiple SDN controllers that have protected access to the internal storage of each other. Hence multi-domain networks can be controlled in a centralized, yet multi-tenanted manner, i.e. without sharing the single controller. This offers multi-tenancy and tenant isolation in the CPS networks which typically have to share the network for the data and control flows unlike the traditional data center networks that can have dedicated bandwidth for each. Thus the controller farm and the westbound API facilitate the execution and interoperability of various entities in \projectName, coordinated by SDN controllers, legacy network controllers, and the other controllers of physical devices and cyberspace.




\textbf{The Eastbound API} is leveraged by the administrators to configure and manage the controller deployment itself. By offering a restricted access to the tenant space in the internal data store of the controller, sensors and actuators in a sensor network can efficiently collaborate and communicate with one another and with the controller. This produces a \textbf{Software-Defined Sensor Network}, that can control sensor networks and heterogeneous smart devices, extending a controller farm of SDN controllers with lightweight southbound MOM protocols. Equipped with i) the global view of the system from the SDN controller, and ii) scalability of the control plane from the controller farm, the Software-Defined Sensor Network makes flow and process decisions based on the tenant preferences and system policies from the cyberspace application layer.

\subsection{\projectName Core Enablers}
\projectName architecture adheres to the functions and attributes of CPS while not sacrificing the capabilities of CPS and to maintain backward compatibility with existing CPS architectures. This can be articulated in a \textbf{5C level} architecture~\cite{lee2015cyber} in a bottom-up approach: i) Software-Defined Sensor Networks representing the \textbf{Smart Connection Level} stays the core bottom-most element in the 5C level architecture which is responsible for plug \& play of sensor networks. \projectName further leverages the controller farm to offer a teather-free communication for the network. ii) \textbf{Data-to-Information Conversion Level} handles multi-dimensional data analytics. \projectName Software-Defined Service Composition visualizes the analytics as microservices and executes the multi-dimensional data correlation as service compositions. iii) \textbf{Cyber Level} supported by the \projectName modelling sandbox offers a twin representation for the physical devices and their cyber counterpart with identification and memory across time, offering data mining capabilities in the cyber representation for decision making. iv) \textbf{Cognition Level} targets the human aspects with modelling, simulation, and visual aspects of CPS. The Software-Defined Simulations enable integrated visualization and synthesis for the Cognition Level. and v) \textbf{Configuration Level} offers self-configuration and adjustment for resilience, optimization, and healing capabilities as the top-most layer of the architecture.

\subsubsection{\textbf{Software-Defined Service Composition}}
CPS networks share the data and control flows over the same network bandwidth, despite the heterogeneity in data flow of various CPS and devices. Hence in order to isolate the bandwidth allocation, the controller farm is leveraged to offer a tenant-aware virtual network allocation. This provides differentiated QoS for various applications and devices sharing the network. The execution is broken into sub executions to enable parallel and independent executions. This Software-Defined Service Composition enables executions as microservices in the control plane. Leveraging the potential multiple alternative execution paths that exist in the devices' execution path and those that are enabled by the virtual tenant network allocation, Software-Defined Service Composition enables breaking down complex computations into distributed service executions that can be executed in parallel in controller and CPS devices' firmware with differentiated priority and control.

Through a common API and an SDN-based approach, \projectName Software-Defined Service Composition enables web services to be composed through various distributed execution paradigms such as MapReduce~\cite{dean2008mapreduce} and Dryad~\cite{isard2007dryad}, in addition to the traditional web services engines to fit the requirements of the CPS. It further allows the services detection and execution to be dynamic, to balance the load across various services nodes. It does so by leveraging the network load information readily available to the SDN controller, as well as the service-level information such as requests on the fly and the requests in the queue that are available to the web services engine, and the services deployment information available to the web services registry.

\subsubsection{\textbf{Modelling Sandbox}}
The \projectName modelling sandbox offers modelling and orchestrating capabilities, thus using the controller as a sandbox in modelling the complex CPS in real world. \textbf{Software-Defined Simulations} bring the simulations of SDN systems close to the systems that they model, where the system being simulated is separated from the simulated application logic. Following a software-defined approach, Software-Defined Simulation models and continuously and iteratively designs the CPS. Thus the simulation in cyberspace will be closer to the execution in the cyber-physical deployment. The modelling sandbox further offers dynamic management capabilities to heterogeneous systems by providing a software-defined approach to orchestrate various stages of development, from simulations, emulations, to physical deployments.

\subsection{Resilience in \projectName}
Ensuring resilience in CPS is a primary goal of \projectName. \projectName attempts to leverage the global knowledge of the entire CPS network to ensure that the elements of the connected CPS are efficiently leveraged in ensuring correct and high performance execution.

Computation power is often rare at the physical location to perform complex computations. Hence, computation-intensive algorithms of the physical devices is delegated to the cyberspace and executed as a composition of microservices, choosing virtual execution spaces in the controller environment. The microservice-based execution avoids repeated computation efforts. The data flow goes through various intermediaries in a traditional workflow. The workflows can be sent through the potential alternatives to ensure load balancing and fair resource utilization. The availability and readiness of redundancy in execution alternatives enables workflows to be executed in a distributed and parallel manner when possible.

\begin{figure}[ht]
	\begin{center}
		\resizebox{\columnwidth}{!}{
			\includegraphics[width=\textwidth]{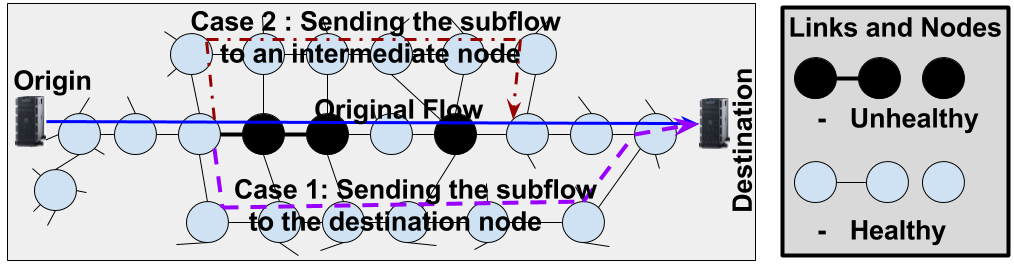}
		}
				\vspace{-1.5em}
	\end{center}
	\caption{Execution as a Service Composition and Alternative Execution Paths}
	\label{fig:action}
\end{figure}

Figure~\ref{fig:action} models a wireframe of the underlying system of CPS with data flow between two smart devices, with multiple potential paths. The origin and destination nodes are the start and the end nodes of a communication caused by a distributed computation. In a data center network, these nodes are hosts or servers, while the intermediate nodes are traditionally switches that connect the large underlying network. However, due to the heterogeneous nature of CPS, origin and/or destination can be smart mobile devices or virtual execution spaces in the controller, while intermediate and/or destination nodes can be surrogate nodes such as computer servers. Without sacrificing the details, \projectName views this as a connected network.

In addition, the path redundancy makes CPS fault-tolerant and ready to handle unexpected failures and congestion. With the dynamic traffic of network flows, a few service or network nodes and links may become congested. Moreover, some nodes may be prone to failures. \projectName attempts to identify the congested, malfunctioning, or malicious nodes and links (that are highlighted and differentiated as unhealthy in Figure~\ref{fig:action} for the ease of reference) through its southbound API.

When an intermediary is identified as failed or slow, \projectName enforces a partial redundancy in the data flows to ensure correctness and end-to-end delivery. \projectName approach creates subflows by diverting or cloning parts of the flows, and sends them towards a node known as the clone destination. In case 1, the clone destination is same as the original destination. However, case 2 has a clone destination that differs from the original. Here the cloned subflow is sent towards an intermediate node on the original path connecting the origin and destination. The flow is recomposed afterwards. The case 2 approach minimizes unnecessary redundancy when it is possible to recompose the flow at the clone destination or an intermediate node. When such a recompose of flows is impossible at an intermediate node due to the technical difficulties or due to the nature of the congestion or network failure itself, the flow is eventually recomposed when it reaches the destination host as in the case 1.


\subsection{Security in \projectName} 
It is essential to secure the controller in \projectName for a correct execution, as an unprotected controller will become a vulnerability on its own. General researches on improving the SDN security are and will be relevant and applicable here, with further extensions for the southbound API for the CPS.

The centralized control avoids the potentials for a network segmentation. Thus, with the global knowledge of the CPS, the \projectName controller mitigates the risks of resource scarcity or external attacks in the intermediate nodes in the underlying network and system. Moreover, the awareness of the application and network enables the controller to differentiate the quality of service (QoS) offered to the tenant applications based on the importance or service-level agreements (SLA).

Nevertheless, distributed fault-tolerance and recovery upon system and network failures are handled efficiently using the controller as a centralized arbiter in the network. As reported for the traditional networks, threats on confidentiality, integrity, availability, and consistency are inherent to the network, and are not introduced by SDN itself~\cite{schehlmann2014blessing}. The vulnerability in privacy due to the co-existence and shared space of tenants, and issues in scale are caused by poor implementation than the design of SDN. \projectName avoids these through the highly available, multi-tenanted, federated controller deployment, designed as the controller farm.

\section{Current Prototype}
\label{sec:impl}

We prototyped \projectName with OpenDaylight~\cite{medved2014opendaylight} Beryllium as the core SDN controller, Oracle Java 1.8.0 as the programming language, and ActiveMQ 5.14.2~\cite{snyder2011activemq} as the message broker of MOM protocols.

\subsection{Modelling and Scaling CPS with \projectName}
The scale and complexity of the CPS increase due to either the larger number of devices and components, or their heterogeneity. Typically, the controller is the element with the highest processing power in the \projectName ecosystem. It manages the communication and coordination across all the entities, including the CPS, humans, and the tenant applications. The federated controller deployment ensures smooth scaling and decision making in the large-scale execution environments. As the controller itself is multi-tenanted with protected access to multiple domains or tenant spaces, management and orchestration of the intelligent agents and their data in the cyberspace are handled seamlessly with scale.

Through Software-Defined Simulations, the designed systems are initially modelled as simulations that are still coordinated by the centralized controller in the same way the physical system that it models is coordinated. Hence, the simulations function as a virtual proxy for the system that is being designed. The systems are in practice implemented once in simulation, and then in physical deployment, reusing the same single effort, having controller as a unified execution space. As the modelling sandbox functions as a controlled modelling space of the designed CPS, unpredictability of the execution environment is significantly reduced.

\begin{figure}[ht]
	\begin{center}
		\resizebox{\columnwidth}{!}{
			\includegraphics[width=\textwidth]{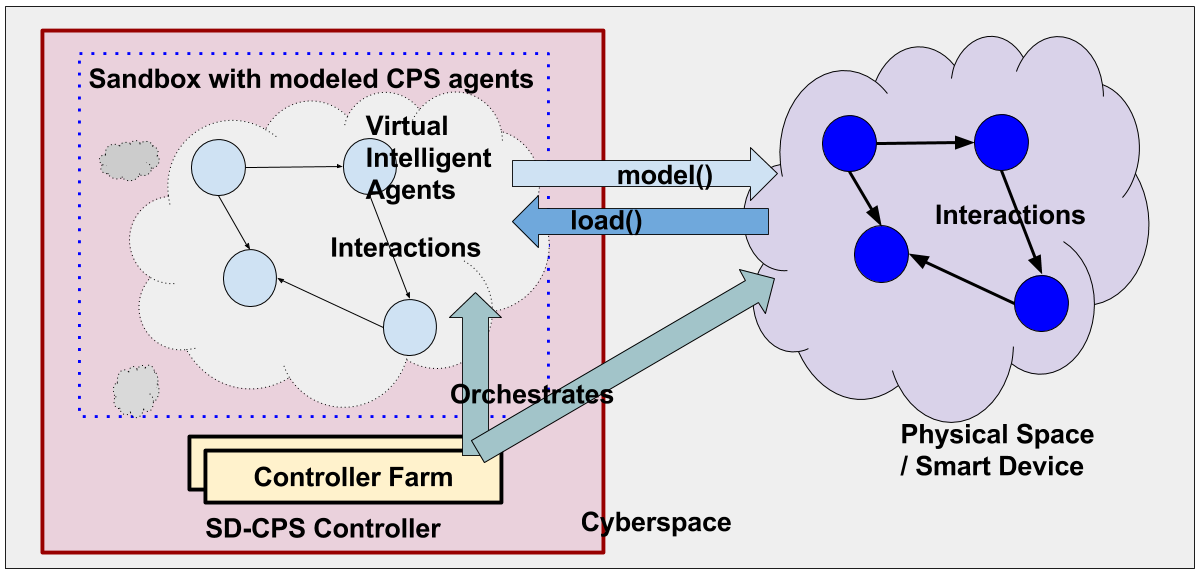}
		}
				\vspace{-2em}
	\end{center}
	\caption{Modelling with \projectName Approach}
				\vspace{-1em}
	\label{fig:sim}
\end{figure}

Figure~\ref{fig:sim} represents how the systems are modelled in the sandbox environment of \projectName controller. The controller farm of \projectName orchestrates both the physical systems and their simulated counterparts in the cyberspace. With a one-to-one mapping between the simulated virtual intelligent agents and interdependent components of the physical system, the interactions are modelled and closely monitored in the controlled sandbox environment before the decisions are loaded into the physical space.

				\vspace{-1em}

The model follows the Software-Defined Simulations and orchestration approach, and attempts to minimize the code duplication by executing the real code from the controller, instead of having a simulation or model running custom code independent of the real execution. As the controller is developed in a high-level language such as Java, \projectName enables deployment of custom applications as controller plugins to alter or reprogram the behaviour of CPS. The physical system loads the decisions from the cyberspace. A multi-tenanted execution space ensures modelling of multiple CPS in parallel.

\subsection{Implementation Details}
\projectName extends and leverages our previous work as the core enablers of the software-defined approach for CPS.

\paragraph*{\textbf{Smart Connection and Data-to-Information Conversion Levels}} 
CHIEF~\cite{7527806} designs the controller farm, a federated deployment of SDN controllers, to manage scalable multi-domain cloud networks. Initially designed for community network clouds, CHIEF was exploited as the \projectName controller farm for any large scale network composed of multiple tenants with heterogeneous devices and access. In addition to the network management, CHIEF offers auxiliary services such as throttling and network monitoring through its event-based extended SDN architecture. \projectName extends Mayan~\cite{kathiravelu2016building} to offer Software-Defined Service Composition for microservices representing the CPS executions. Cassowary~\cite{kathiravelu2015cassowary} designs Software-Defined Sensor Networks for smart buildings leveraging SDN and MOM protocols. We extend Cassowary to facilitate a wider adoption of SDN with loose coupling to the underlying network or SDN switches.

\paragraph*{\textbf{Cyber and Cognition Levels}} 
SDNSim~\cite{kathiravelu2016software} offers Software-Defined Simulations. Built on top of SDNSim, SENDIM~\cite{kathiravelu2016sendim} enables systems to be designed and deployed seamlessly across various realizations and deployments. Originally developed for cloud and data centers, SENDIM is extended for CPS, IoT, or any software-defined systems and networks, as the modelling sandbox of \projectName.

\paragraph*{\textbf{Configuration Level}} 
Core configuration data is stored in the controller by exposing its persistent in-memory data store through the REST and MOM protocol implementations of \projectName northbound API. The data store of \projectName extends the OpenDaylight controller data tree.

\subsection{Feasibility Assessment}
Through a few simulations and microbenchmarks, we demonstrated that \projectName increases the potential scale of the CPS. \projectName controller performance was increased through the deployment of controller farm~\cite{7527806}. A near-linear performance growth with the number of controller instances up to a maximum value followed by a near-logarithmic growth was observed~\cite{kathiravelu2016building}. The reduced performance gain is due to idling controllers for each service execution. Hence, the performance gain depends on the problem size and its distribution as services in service composition. Furthermore, the modelling sandbox reduces the time in modelling as it offers a dual reality of cyber-physical spaces for simulations and designs.

\section{Related Work}
\label{sec:related_work}
Use cases of SDN have been steadily spanning beyond the traditional networks, from sensor networks to smart buildings. 

\paragraph*{\textbf{SDN and OpenFlow}} Wireless Sensor Networks (WSN)~\cite{romer2004design} have the requirement to be context-aware. They need to handle a larger control traffic due to their dynamic nature compared to data center networks, while having to share the bandwidth among control and data traffic. Sensor OpenFlow (SOF)~\cite{luo2012sensor} identifies the benefits of a Software-Defined WSN, leveraging SDN for WSN. SOF increases manageability of WSN and adapts to policy changes of wireless networks and mobile devices.

Albatross~\cite{leners2015taming} discusses the challenges faced by distributed systems, and aims to mitigate them by leveraging SDN. The challenges such as split-brain scenarios and violations in consistency and availability that are addressed by Albatross are relevant for CPS too. However, while CPS is a distributed system, it has its own peculiar challenges due to its diverse nature in implementation and devices as we discussed earlier.

\paragraph*{\textbf{Smart Environments and CPS}}

Software-Defined Environment (SDE)~\cite{dixon2014software,li2014software} focuses on factors such as i) resource abstraction based on capability, ii) workload abstraction and definition based on policies, goals, and, business/mission objectives, iii) workload orchestration, and iv) continuous mapping and optimization of workload and the available resources. SDN controller and physical and virtual SDN switches remain the heart of SDE. The control of compute, network, and storage is built atop a virtualized network.

Software-Defined Buildings (SDB)~\cite{dawson2012energy} envision a Building Operating System (BOS) which functions as a sandbox environment for various device firmwares to run as applications atop it. The BOS spans across multiple buildings in a campus, than confining itself to a single building. SDB and SDE architectures can be extended for CPS. However, they cannot cater for CPS by themselves due to the variety and heterogeneity in the architecture and requirements of CPS compared to the environments controlled by SDB and SDE.

\paragraph*{\textbf{Software-Defined Internet of Things (SDIoT)}}
SDIoT~\cite{jararweh2015sdiot} proposes a software-defined architecture for IoT devices by handling the security~\cite{al2015sdsecurity}, storage~\cite{darabseh2015sdstorage}, and network aspects in a software-defined approach. SDIoT proposes an IoT controller composed of controllers of software-defined networking, storage, security, and others. This controller operates as an orchestrating middleware between the data-as-a-service layer consists of end user applications, and the physical layer consists of the database pool and sensor networks.

Multinetwork INformation Architecture (MINA) self-observing and adaptive middleware~\cite{qin2014mina} has been extended with a layered SDN controller to implement a controller architecture for IoT~\cite{qin2014software}. Various research and enterprise use cases are proposed and implemented, including SDIoT for smart urban sensing~\cite{liu2015software}, and end-to-end service network orchestration~\cite{vilalta2016end}. While sharing similarities with IoT, CPS is set to address a larger set of problems with more focus on ground issues on interoperability of cyber and physical spaces and dimensions in a CPS. Hence, \projectName differs in scope to that of SDN for IoT researches such as SDIoT, though they share similar motivation.

				\vspace{-0.4em}

\section{Conclusion and Future Work}
\label{sec:conclusion}
\balance
In this paper we presented \projectName, an approach and architecture that aims to mitigate the application and design challenges faced by CPS. \projectName leverages the SDN switches and controllers when available, while employing an approach motivated by SDN even during the absence of SDN switches. Hence it remains compatible with and applicable to existing CPS deployments that do not have SDN. \projectName opens up many research avenues on envisioning and improving SDN for CPS architectures and evaluating implementation alternatives. As a future work, the proposed approach should be deployed in various CPS and tested for its efficiency in addressing the identified shortcomings.
	\vspace{1em}

\scriptsize{
\textit{\textbf{Acknowledgements:}} This work was supported by national funds through Funda\c{c}\~{a}o para a Ci\^{e}ncia e a
Tecnologia with reference UID/CEC/50021/2013 and a PhD grant offered by the Erasmus Mundus Joint Doctorate in Distributed Computing (EMJD-DC).
}

\bibliography{references}

\end{document}